# Speed-Up Ramp-Counter ADC Using Locality Principle: A Systematic Analysis

HoseinAli Jafari Abeshoori
School of Electrical Engineering
Iran University of Science and Technology
Tehran, Iran
hoseinalijafari@elec.iust.ac.ir
https://orcid.org/0000-0001-6928-6666

Seyed Saeed Azadfar
alumni student
*Electrical and Computer Engineering Faculty*
Yazd University
Yazd, Iran
mailto:saeedazadfar@yahoo.com

*Abstract*—This work inspects the parameters that affect the conversion rate of a ramp-counter ADC and propose two small systematical changes to reduce the time needed to complete a single analog to digital conversion by use of the locality principle point of view. System full architecture is modeled and simulated by Matlab script and the results show up to 97% reduction in conversion time for an electro-cardiograph record.

*Keywords—ADC, ramp, counter, locality, speed*

## I. Introduction

Analog to Digital Converter (ADC) is a crucial part of most of the electronic systems and also a bottleneck for communicating between analog and digital circuits. Some types of ADCs use different methods to convert a continuous analog signal into a digital number. Two important parameters in analog to digital conversion are accuracy and conversion rate. A finer conversion with less error and loss of information usually comes with a lower conversion rate in a specific ADC and also a faster converter may prevent extra loss of input samples and reach maximum available accuracy. Therefore, the conversion rate has a very important effect on the output provided by ADC.

For low-power[1]–[6] and area-limited[7], [8] applications without the need for a comparably high sampling rate, a good choice is Ramp ADC[9], [10]. Ramp ADC is introduced in some different topologies such as single and dual slop mode integrating time to amplitude converter[11], [12]. Another type is ramp-counter ADC which uses a counting register beside a simple DAC[13] instead of a timer circuit with a fixed-slope ramp signal generator[14]. In this topology (Fig. 1) for each conversion, a new sample of the input analog signal is inserted and held at the first step. After that, a binary counting register starts to count-up from zero state, and its value is converted to analog amplitude for comparison with the input signal. When the counted value gets enough closed to the input signal, its value is mentioned to be the digital converted value of the input signal and put to the output of the ADC. By starting the next sampling cycle, the counter register value is cleared to zero for counting up again.

In computer science, the principle "locality of reference" points to the tendency of the processor to use the same data that it has accessed recently [15-20]. The locality is referring to the temporal and spatial position of a data set. In other words, when a memory location is used, it is more probable for its near locations to be used in the next processing step. With this in mind, a quick look at analog signals gives that analog signal is continuous in time and domain and therefore it may follow the principle of locality.

By applying the locality principle on a sampled analog signal in the Sample and Hold circuit, for its next sampled value it is expected to be near the last one than to the farther values. This point of view is described, theoretically analyzed, and formulated in section II. Section III proposed a counter ADC simulated by Matlab scripting and the results in both usual and improved architecture are presented. Finally, the conclusion and the suggested works for future studies are gathered in section IV.

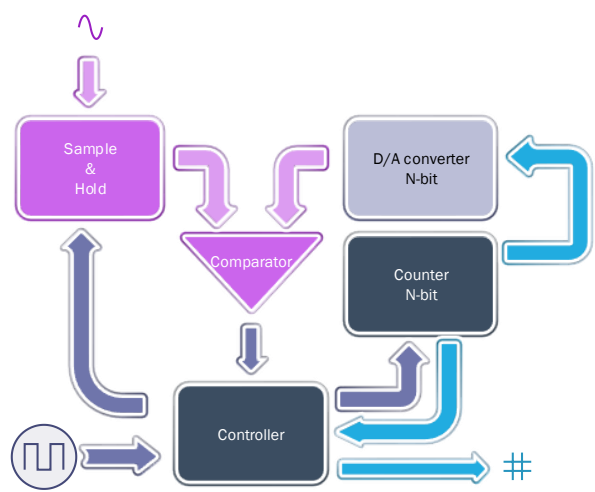

Fig. 1. the internal structure of a typical counter ADC. Control, digital and analog signals, and circuits are sketched in different colors.

## II. Method

The improvement in the architecture of the ADC controller unit includes two changes. (1) The up-counting register must change to an up/down counting one. Also (2) the digital controller of the ADC should be changed so that it could determine the direction of the counting based on the difference between the newly inserted analog value from the S&H circuit and the analog output of the internal DAC.

When an analog signal is inserted by Sample and Hold to be converted, the control unit decides the direction of counting. Since the last binary value of the register from the latest conversion is not cleared, its analog scaled value came from the output of the DAC is compared with the input analog signal. If the analog input signal is greater than the DAC output, the conversion starts in an up-counting direction. In the case of greater DAC output compared with the input analog signal of S&H, counting the register in a downward direction reduces the difference between the DAC output and the reference signal of the S&H.

In a normal counter-ADC with register clearing after each conversion, the time to complete a conversion depends on signal value and parameters of the ADC which is expressed in (1) where S stands for analog signal amplitude and the reference signal is the maximum allowed input signal of the ADC that can be converted linearly before it gets saturated.

$$T_C = \frac{\left\lfloor \frac{S_{S\&H}}{S_{REF}/(2^{BIT}-1)} \right\rfloor}{T_{count}} + T_S + T_F \quad (1)$$

TS is the Sampling time needed for the Sample and Hold circuit to take a sample value from the input analog signal, hold and insert it for converting to a digital binary number. When the counting got enough close to the target input sample, a little Finalizing time is needed for the control unit to put out the calculated binary number as the output of the ADC. Both $T_S$ and $T_F$ in a counter ADC is considered to be almost constant for each conversion and a few times (M, N) of base clock pulse period of the ADC. The counting time gaps of the register are also constant and could be noticed as a factor (K) of the clock pulse period. Since the number of the output bits for a normal ADC is usually greater than 4, so the denominator of the reference signal $S_{REF}$ can be approximated to $2^{BIT}$. Reforming the equation (1) simplifies it to (2):

$$T_{Ct} \approx \left\lfloor \frac{S_{S\&H}}{S_{REF}} 2^{BIT} \right\rfloor (K.f_{CLK}) + M.f_{CLK} + N.f_{CLK} \quad (2)$$

It is understood from (2) that conversion time $T_C$ is a factor of the base clock pulse period. By compressing the constant values to the L, it is formulated in (3) where $T_{count}$ is assumed to be equal to $T_{CLK}$ which means the register counts up or down in each clock pulse of the controller.

$$T_{Ct} = f_{CLK}\left(\left\lfloor \frac{S_{S\&H}}{S_{REF}} 2^{BIT} \right\rfloor + L\right) \quad (3)$$

In the same manner for the proposed architecture of fig. (2), the conversion time is calculated as (4). It can be assumed that in (3) the $S_{LAST}$ is zero. This type of formulation is very useful for comprehension and comparison of the speed of the typical and proposed architecture.

$$T_{Cp} = \left\lfloor \frac{|S_{S\&H} - S_{LAST}|}{S_{REF}} 2^{BIT} \right\rfloor (K.f_{CLK}) + M.f_{CLK} + N.f_{CLK} \quad (4)$$

The same assumption for the ADC parameters results in (5) where the register value lasts after completion of the previous conversion is named $S_{LAST}$.

$$T_{Cp} = f_{CLK}\left(\left\lfloor \frac{|S_{S\&H} - S_{LAST}|}{S_{REF}} 2^{BIT} \right\rfloor + L\right) \quad (5)$$

To have an estimation of $S_{LAST}$, it should be mentioned that $S_{LAST}$ is the previous value of $S_{S\&H}$ before starting the new conversion cycle. Therefore their difference can be calculated as the mean slope of the analog signal in the time gap between the last two taken samples Δt.

$$S_{S\&H} = S_{LAST} + \langle \tfrac{d}{dt} S_{input} \rangle \times \Delta t \quad (6)$$

Since the usual frequencies of the clock pulse are much more than the bandwidth of the analog signal, which is implied by Nyquist sampling rate, the derivative of the $S_{input}$ has very small changes in the time gap of Δt. so the averaging could be neglected and replaced by its instantaneous derivative. Also, it is obvious that the time gap between the two samples is the conversion time of the last sample (7).

$$\Delta t = T_{C_{(n-1)}} \quad (7)$$

By substituting the equations (6) and (7) in (5), the conversion time of the proposed architecture is (8) which is a recursive function of $T_{C(proposed)}$ recent values. $T_{C(n-1)}$ is needed for calculating the conversion time of the next sample $T_{Cp(n)}$.

$$T_{Cp_{(n)}} = f_{CLK}\left(\left\lfloor \left|\tfrac{d}{dt} S_{input}\right| \times T_{Cp_{(n-1)}} \times \frac{2^{BIT}}{S_{REF}} \right\rfloor + L\right) \quad (8)$$

For both types, typical and proposed architectures of the counter ADC, the least possible conversion time $T_C$ is limited by overhead time of (9). In a typical architecture, $T_C$ increases by enlarging the input analog signal to higher values. But in the proposed architecture, $T_C$ is independent of input analog signal amplitude. From (8) it is apparent that the rate of variation of the input signal is determinant for conversion time. Since the clock pulse is usually much faster than the input analog signal variations, it is theoretically expected for $T_{C(proposed)}$ to be not much larger than $L \times f_{CLK}$.

$$T_C \geq f_{CLK}.L \quad (9)$$

A reduction factor for the conversion time is introduced in (10). Then substituting of (3) and (8) in (10) results in (11) which can be estimated to (12) using the fact that in a normal condition the average value for an analog input signal is about half of the reference level ($S_{REF}$).

$$r = 1 - \frac{T_{C_{proposed}}}{T_{C_{typical}}} \quad (10)$$

$$r = 1 - \frac{\left\lfloor \langle \left|\tfrac{d}{dt} S_{input}\right| \times T_{C_{(n-1)}} \rangle \times \frac{2^{BIT}}{S_{REF}} \right\rfloor + L}{\left\lfloor \frac{S_{S\&H}}{S_{REF}} 2^{BIT} \right\rfloor + L} \quad (11)$$

$$r \approx 1 - \frac{\left\lfloor \left(\left|\frac{d}{dt}S_{input}\right| \times T_{C_{(n-1)}}\right) \times \frac{2^{BIT}}{S_{REF}}\right\rfloor + L}{2^{BIT-1} + L} \quad (12)$$

Equation (12) shows that the reduction ratio is statistically limited to (13) for fine and fast enough ADCs. When the ADC has a clock pulse frequency of much higher than input analog signal swinging, nominator of (12) approaches to L. for fast and fine ADCs, L is negligible when compared to $2^{BIT-1}$ and the denominator could be estimated to $2^{BIT-1}$.

$$r < 1 - \frac{L}{2^{BIT-1}} \quad (13)$$

To have a measurement from the available working parameters of an ADC, counting the converted samples of the analog input signal is a more feasible way. The number of collected samples $N_S$ in a long enough time frame is inversely related (14) to the average of the conversion-time $T_C$.

$$N_S = \frac{T_1 - T_0}{\langle T_C \rangle} \quad (14)$$

$T_0$ and $T_1$ are the start and stop points of the measuring time frame where the $T_C$ is averaged out. Another useful figure of merit is the speed-up ratio which is defined as (15).

$$Speed\ Up = \frac{N_{S(proposed)}}{N_{S(typical)}} \quad (15)$$

Substituting (14) in (10) puts up the relation between reduction factor and number of collected samples in (16) which can be simplified and formulated as (17) by use of speed-up definition (15) that is presented in percents.

$$r = 1 - \frac{(T_1 - T_0)/N_{S(proposed)}}{(T_1 - T_0)/N_{S(typical)}} \quad (16)$$

$$r_\% = \left(1 - \frac{1}{Speed\ Up}\right) \times 100 \quad (17)$$

### III. SIMULATION RESULTS

The function of a typical and the proposed counter ADC is implemented through Matlab scripting. Four different types of analog signals are applied as input to inspect the reduction factor in most possible cases. The time frame of the input signals is normalized and the Clock pulse frequency is expressed as a factor of the time frame instead using of real values for time and frequency. Allowed input amplitude is considered to be between zero and one without any dimension. For better observability and comparison, the output binary number of the ADC is plotted on the input curve. Quantized output levels are also normalized and divided into $2^{BIT}$ levels.

Fig. 3 shows the result of the analog to digital conversion for both architectures on a common figure with the same Bit count and clock frequency. The input analog signal, its digital converted values by typical and proposed architectures are sketched in green, red, and black respectively. The input signal is a sine wave sampled and converted for a single period by a 6 BITs ADC in different clock pulse speeds.

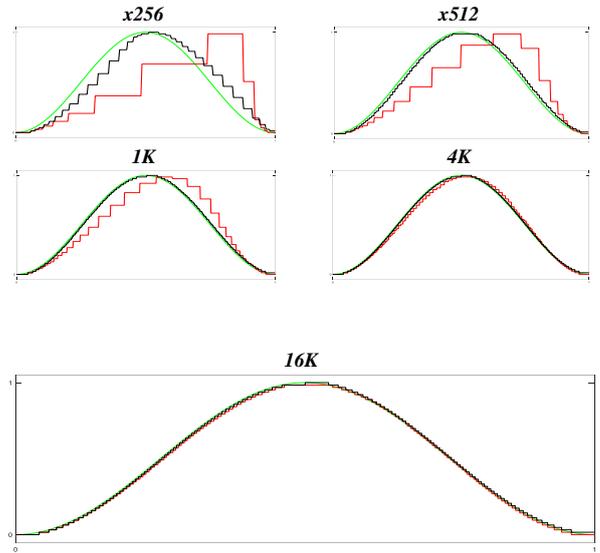

Fig. 2. Sampling and converting of a single sine wave(green) for a single cycle with both typical(red) and proposed(black) architectures

Conversion results for Typical and proposed architecture reach its acceptable target for clock frequencies of 16000 and 512 times the sine wave respectively. An output with more BIT counts regularly needs more times or clock periods to be calculated and consequently needs higher clock frequency to reach its best point compared to inputs signal bandwidth top boundary. It is obvious that the proposed architecture reduced the minimum allowed clock frequency of the ADC that guaranty minimum acceptable error or distortion from the input signal.

To find the reduction factor which is a statistical value, ADC is tested for some different kinds of input analog signals. Choosing these types based on the parameters that appeared in the theoretical analysis may affect the result. A 10% offset DC signal has no variation and led to a maximum conversion rate. An Electrocardiography (ECG) record is used as a fast varying signal. An exponential ($e^t$) ramping up input inserted as a slow varying full-scale range signal and a sine wave for moderate-speed with large-swing one. The configuration of the ADC is 8 bits of binary output with 256K clock pulses for the specified unit time frame of the fig. (4).

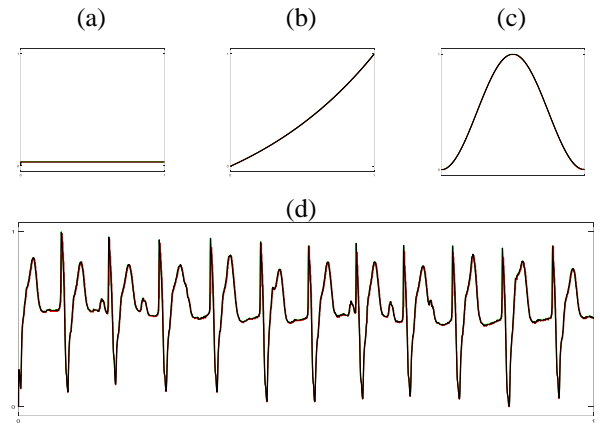

Fig. 3. Sampling and converting input(green) signals of type (a)DC, (b)exponential, (c)Sine-wave and (d)ECG for a fixed unit time-frame with both typical(red) and proposed(black) architectures

Because of the high enough number of bits and system clock pulse speed, signal tracking and conversion are very fine with a little lag and error. But the amount of samples taken from the input and converted to digital number differs for two architectures. The number of samples collected for each signal is reported in the table.1 separately for both architectures with calculated speed-up and reduction factor in percent.

TABLE I. TABLE TYPE STYLES

| Input Signal | Collected Samples | | Calculated F.O.M. | |
| --- | --- | --- | --- | --- |
| | Typical | Proposed | Speed-up | Reduction factor |
| DC | 21845 | 87378 | 3.99 | 75% |
| Sine | 10,064 | 87,212 | 8.66 | 88% |
| Exponential | 6,520 | 87,296 | 13.4 | 92% |
| ECG | 2,378 | 85,128 | 35.8 | 97% |

Fortunately, the top most gotten value for reduction factor found in case of ECG record which is an example of a real application for use of ADC in biomedical instruments. Looking at the columns of the collected samples count reveals that in the proposed architecture there is less variation in conversion rate compared with the typical one. The results are in good accordance with theoretical expectations.

IV. CONCLUSION

By adding the ability to counting in up and down direction, a little complexity is added to the counter and controller units but the frequency of the clock pulse needed for converting the same input signal is much less and its power consumption would be less by far. On the other hand with having the same technology and clock pulse speed, by use of the proposed architecture, signals of higher bandwidth could be sampled and converted. Improvements for input signal bandwidth, clock pulse speed, and power consumption depends on the aspects and specifications of the input analog signal.